\documentclass[aps,prl,twocolumn,showpacs,superscriptaddress,10p,longbibliography]{revtex4-1}
\usepackage{graphicx}
\usepackage{epsfig}
\usepackage{dcolumn}
\usepackage{amsmath}

\usepackage[sort&compress]{natbib}

\usepackage{bm}
\usepackage{bbm}
\usepackage{ulem}
\usepackage[colorlinks, citecolor=red]{hyperref}
\usepackage{mathtools}
\usepackage{color}
\usepackage{soul}
\usepackage{centernot}
\usepackage[up]{subfigure}
\usepackage{dsfont}	
\usepackage{relsize}
\usepackage{epstopdf}
\usepackage{mathrsfs}
\usepackage{braket}

\newsavebox{\graphicsbox}

\DeclareGraphicsExtensions{.pdf,.png,.jpg}
 
\newcommand{\curl}{\nabla \times}
\newcommand{\dive}{\nabla \cdot}

\newcommand{\be}{\begin{equation}}
\newcommand{\ee}{\end{equation}}
\newcommand{\ba}{\begin{eqnarray}}
\newcommand{\ea}{\end{eqnarray}}

\newcommand{\vect}[1]{\mathbf{#1}}

\newcommand{\bas}{\begin{eqnarray*}}
\newcommand{\eas}{\end{eqnarray*}}
\begin{document}
\title{Measurement and Memory in the Periodically Driven Complex Ginzburg-Landau equation
}
   \author{T. Mithun}
   \affiliation{Department of Mathematics and Statistics, University of Massachusetts, Amherst, MA 01003-4515, USA}
   
   \author{P. G. Kevrekidis}
   \affiliation{Department of Mathematics and Statistics, University of Massachusetts, Amherst, MA 01003-4515, USA}

\author{A. Saxena}
   \affiliation{Center for Nonlinear Studies and Theoretical Division,
Los Alamos National Laboratory, Los Alamos, New Mexico 87545, USA}
  
\author{A.R. Bishop}
   \affiliation{Center for Nonlinear Studies and Theoretical Division,
Los Alamos National Laboratory, Los Alamos, New Mexico 87545, USA}

\begin{abstract}
In the present work we illustrate that classical but nonlinear systems 
may possess features reminiscent of quantum ones, such as memory, 
upon suitable external perturbation. As our prototypical example,
we use the two-dimensional complex Ginzburg-Landau equation in its vortex glass regime. We impose an external drive as a perturbation mimicking a quantum measurement protocol, with a given 
``measurement rate'' (the rate of repetition of the drive) and ``mixing 
rate'' (characterized by the intensity of the drive). Using a variety of measures, we find that the system may or may not retain its
coherence, statistically retrieving its 
original glass state, depending on the strength and periodicity of the 
perturbing field. 
The corresponding parametric regimes and
the associated energy cascade
mechanisms involving the dynamics of vortex waveforms and domain boundaries are discussed. 
 \end{abstract}
\maketitle
\section{Introduction}
This report examines the property of memory under
periodic driving in a classical nonlinear system, namely
the complex Ginzburg-Landau (CGL) equation ~\cite{RevModPhys.74.99}. Our aim
is to demonstrate properties of memory and coherence,
following measurement probes. This is an interesting topic in its own right but also illustrates classical precursors to phenomena in quantum information contexts. The interplay between nonlinearity and quantum mechanics
is an important ingredient in the fascinating but
complicated issue of quantum-classical correspondence \cite{pang2005quantum,bolivar2013quantum}. The intense recent focus on quantum information science
and technology and associated potential physical sources
of qubits provides new impetus for this topic.
Clearly some properties are purely quantum in their nature \cite{kotler2021direct,de2021quantum}. However, other important features have quantum
parallels. Measurement rates and decoherence are major
concerns for quantum information devices. Classical
analogs appear in glassy and disordered systems and more
generally in appropriate classes of nonlinear equations ~\cite{PhysRevA.65.032321,PhysRevA.62.012307}, where analogs of coherence and decoherence, entangled
states ~\cite{PhysRevA.65.032321,spreeuw1998classical},
chimeric patterns \cite{yao2013robustness}, frustration, and complexity \cite{guastello2008chaos,guastello2009introduction} are very rich. 
Another prominent example where such analogies have been intensely
pursued over the past decade concerns the emerging field of the so-called
pilot-wave hydrodynamics~\cite{bush1,bush2}.

Our purpose here is to explore a nonlinear example of “measurement” and “memory” from this perspective. Many nonlinear equations can be used to explore these parallels, such as the nonlinear Schr\"{o}dinger (NLS) model~\cite{sulem,ablowitz,Kevrekidis2015}, the sine-Gordon equation \cite{cuevas2014sine}, coupled double-well potentials \cite{christoffel1981quantum}, etc. Indeed, nonlinearity often arises as a semi-classical approximation to quantum many-body systems (e.g., in BCS superconductivity \cite{tinkham2004introduction}, charge-density waves, Josephson junction equations \cite{pino2016nonergodic,PhysRevLett.122.054102}, or the Gross-Pitaevskii equation (GPE)~\cite{pethick,stringari}). Typically, nonlinear equations are the result of slaving among coupled linear fields, and their properties can be controlled by a variety of external drives.
In fact, the NLS  has been studied 
as a result of a quantum feedback process upon identical quantum systems subject to weak
measurements~\cite{PhysRevA.62.012307},
while the GPE emerges as a result of a
mean-field approximation of cold, dilute atomic quantum
gases. These lines of analogy between quantum systems
and nonlinear classical ones have led to the study of classical analogues of entanglement
 \cite{PhysRevA.65.032321,spreeuw1998classical} and decoherence \cite{gong2003quantum}, the latter including a nanomechanical resonator \cite{maillet2016classical}. 

We have chosen here to consider the two-dimensional (2D)
CGL, as a prototypical example of such connections \cite{RevModPhys.74.99}. 
The CGL is interesting because it represents a {\it generic} amplitude model in the
presence of slow variation and weak nonlinearity, corresponding to 
a normal form-type partial differential equation near a primary bifurcation of a homogeneous state~\cite{RevModPhys.74.99}. At the same time, it exhibits secondary excitations in the form of topological coherent structures (domain walls, solitons and vortices), contains explicit self-consistent dissipation, and has been extensively used to successfully model phenomena in a number of physical systems,
extending from superconductivity and superfluidity to liquid crystals, Rayleigh-Benard convection 
and Bose-Einstein condensates~\cite{Bodenschatz_annurev,RevModPhys.74.99}.
In addition, it provides an appropriate vehicle to explore our  interests here as it possesses certain instabilities and coherent structures, as well as highly nontrivial metastable states, such as a vortex glass~\cite{bohr,PhysRevE.65.016122} that we will explore in what follows. As we will see, the topological excitations and their interactions are a key source of (space and time) multiscale patterns and transitions, including freezing and stretched exponential relaxation, which relate to our interest in entanglement and decoherence.

Here we specifically consider the 2D CGL with a cubic
nonlinearity and an external periodic driving coupled to
the field amplitude. It is worthwhile to note here that periodically-forced variants
of the CGL equation have been argued as offering a generic model
of parametrically forced systems such as the Faraday wave experiment~\cite{alastair}.
In this context of hydrodynamics, indeed, there exist well-documented
examples of matching the CGL and its coefficients to 
concrete hydrodynamic experiments~\cite{LEGAL2003114}.
By periodically driving the CGL equation,
we induce an analogue of an external perturbation akin to a
periodic measurement process on a quantum system.
In this driven system we can anticipate a  ``phase diagram" relevant to a classical analog of decoherence: 
one axis is akin to the ``measurement rate" and the other axis represents the 
effective ``mixing rate", induced by the drive. We might anticipate a demarcation curve, separating regions of maintaining and losing coherence, i.e., ones
preserving the memory of the original state and ones losing it. Here, the measurement rate would correspond to a field pulse applied to the system periodically. This will mix the states of the system. After a pulse, if the system recovers, the scenario will be analogous to that of coherence in a quantum system. The mixing rate is proportional to the strength of the nonlinearity in the CGL equation. The stronger the nonlinearity, the higher the expected mixing rate. 

As perhaps the simplest case, we apply the driving field
uniformly in space, at a periodic rate in time, and with
periodic boundary conditions. We numerically follow a variety of diagnostics including ones quantifying the induced vorticity, the compressible and
incompressible energy spectra and the cascades that they reveal and use the creation and annihilation of the
vortices as a prototypical illustration of maintaining coherence, or potentially losing memory of the initial condition
of the system. We identify both of these regimes and delineate the transition between them. We believe that
this type of phenomenology may be broadly applicable in other distributed nonlinear dynamical systems and that
this study may pave the way for further efforts to identify features of quantum systems that have non-trivial classical precursors. Indeed, this and related systems could offer interesting additional paradigms to the rapidly developing area of pilot-wave
hydrodynamics~\cite{bush1,bush2}.

Our presentation will be structured as follows. In Sec. II we present the basic features of the model. Section III
encompasses the quench dynamics protocol and associated observables. In section IV, we comment on our numerical
findings, while section V concludes the report offering some perspectives for future work.



 %
\section{The Model: 2D-CGL equation.}\label{model}

We consider the two-dimensional (2D) dimensionless CGL equation in the prototypical form~\cite{RevModPhys.74.99}
\begin{equation} \label{eq:cg1}
\begin{split}
 \frac{\partial}{\partial t}  A 
  &= A+ (1+i \beta) \Delta A -(1+i \alpha) |A|^2 A\\& +A_0 \delta (t- [T_0+l T]),~~l=0,1,2,...
\end{split}
\end{equation}
Here, $A$ is a complex field and $\alpha$ and $\beta$ are real parameters. $A_0$ represents the strength of the external field (associated
with the mixing rate, as discussed above), while $T$ represents the period of the quenching (i.e., the measurement rate)  and $T_0$  
is the initial offset time which allows the system to relax before the quenching starts. For $\alpha = \beta =0$, this equation becomes a
regular CGL~\cite{RevModPhys.74.99}. On the other hand  in the limit $\alpha, \beta \rightarrow \infty$, this equation reduces to the nonlinear Schr\"{o}dinger equation. 

This 2D model admits a variety of coherent structures, including vortices and domain walls \cite{RevModPhys.74.99,bohr,PhysRevE.65.016122,JANIAUD1992269}. The symmetry breaking instabilities of the model are well studied: these include the Benjamin-Feir, Eckhaus, convective, phase instability and absolute 2D instabilities~\cite{PhysRevA.46.R2992}. A general overview of the model 2D findings can be found in \cite{RevModPhys.74.99}. The phase diagram associated with the system suggests that there exist three pattern
formation regimes, namely defect turbulence, phase turbulence and vortex glass, depending on the parameter values of the model \cite{RevModPhys.74.99,PhysRevE.65.016122}. The peculiarity of the vortex glass state is that in this regime the vortices are arranged in cellular structures thought of as frozen states, i.e., very
long lived metastable states~\cite{PhysRevE.65.016122}. It is the latter
frozen type of extremely long-lived attractor of the system
that we find particularly appealing for our explorations of
(nonlinearity induced) coherence vs. decoherence and would like to further explore. In what follows, we will use quench dynamics into this
vortex glass regime and will subsequently perturb the relevant state
using the external drive.

\begin{figure*}[!htbp]
\includegraphics[width = \textwidth]{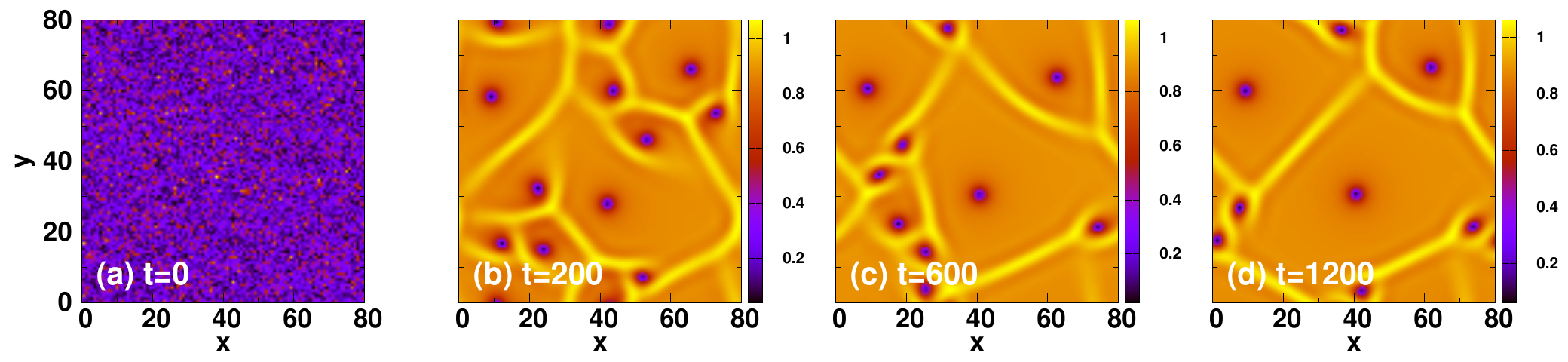} 
\caption{Snapshots of the absolute value of the field ($|A|$) for $\alpha=0.7$ and $\beta=-0.7$ at different times. We consider the state at $t=1200$ as the initial condition for the rest of the numerical studies reported in this work. Movies are available at the link \url{https://www.dropbox.com/sh/65ze3j94l3bcw4v/AADF_UjJeM-EAIgUlno_noiia?dl=0}}
\label{fig:fig1}
\end{figure*}

\subsection{Quench Dynamics}
An absolute instability induces defect pair formation in the model as  demonstrated, e.g., in
\cite{PhysRevA.46.R2992}. This defect pair formation mechanism can be systematically controlled via the nonlinear model parameters $\alpha$ and $\beta$~\cite{bohr}. In this work we fix $\alpha=0.7$ and $\beta=-0.7$. For this set of parameter values, a random initial condition leads to the formation of a vortex glass state as shown in Fig.~\ref{fig:fig1}. For our numerical experiments, we consider a 2D plane of size $L_x \times L_y$, ($L_x =L_y=80$) with $512 \times 512$ grid points.  We solve Eq.~\eqref{eq:cg1} numerically by using an explicit Runge-Kutta algorithm of order 8, namely DOP853~\cite{freelyDOP853,mine-01-03-447}.  We initialized the simulation with $A=0.001*R$, where $R$ represents  random numbers drawn from the uniform distribution $[-0.5,0.5]$ as shown in Fig.~\ref{fig:fig1}(a). Additionally, periodic boundary conditions $A_{i,0} = A_{i,M}$, $A_{i,M+1} = A_{i,1}$, for $1 \le i \le L$ and $A_{0,j} = A_{L,j}$, $A_{L+1,j} = A_{1,j}$, for $1 \le j \le M$ are considered.

The snapshots of field amplitude at different times in Fig.~\ref{fig:fig1}(b-d) show the vortex defects that are surrounded by shock lines \cite{chate1996phase}. They form a long-lived metastable cell structure together. The vortex defects are known to produce spiral waves \cite{aranson1992stability}. In this work we are interested in performing amplitude quenches to excite  the system and observe
its potential return to equilibrium (or absence thereof). We consider the field shown in Fig.~\ref{fig:fig1}(d) at $t= 1200$, as corresponding to $T_0$ in Eq.~(\ref{eq:cg1}), i.e., as the initial condition for the following numerical experiments. This state is a relaxed state in the sense that
no more vortex-antivortex annihilations take place upon further evolution until at least $T=2000$. On the other hand, this is a metastable state in which (slow) vortex motion still persists. Hence, such a state is often referred to  as a glassy state~\cite{PhysRevE.65.016122}.

We now periodically drive the (glassy state of the) system for a fixed amplitude $A_0$ (real variable) and the period $T$. We fix the time periods $T=10$ and $T=15$ for our numerical experiments. These time periods are relatively short as compared to the time required for the system to relax. As a result, a periodic drive takes the system  out of equilibrium by changing its total energy  and we then
observe the subsequent response of the system to both a single 
excitation, but also importantly to a periodic sequence of such
excitations.
 
 \subsection{Observables}
 We consider the normalized distribution $P_{i}=P(n_i)$, where $P(n_i)$ is the probability distribution of amplitudes (PDA) $n_i=|A_i|$. 
 For the complex CGL system,  we first define the observable Shannon entropy
  \begin{equation}
S_v =-\sum_{i}\Big[P_{i}\log(P_{i})\Big].
  \end{equation}
 as a standard information-theoretic diagnostic of the system.
  
Moreover, since the topoplogical excitations in 2D of the CGL model are vortices, we additionally consider the following observables for the measurements. We measure the absolute value of the winding number \cite{Akhavan:2020} 
    \begin{equation}
\Gamma =\frac{1}{2\pi}\int |\omega|dx dy ,
  \end{equation}
   where $\omega$ is the vorticity. The vorticity is related to the  velocity field $\vect{v}$ as $\bf{\omega}=\nabla \times \bf{v}$. This diagnostic counts the total number of vortices in the system.
   The change in this number will be used to assess the potential
   loss of memory through the creation of additional
   excitations and the departure from the previously reached glassy
   state, upon introducing the relevant perturbation. 
   
   Further, in order to follow the energy distribution in the system we separate the field into compressible and incompressible parts. The term $| \nabla A|^2 $ [i.e., the energetic contribution 
   associated with the Laplacian in Eq.~\eqref{eq:cg1}] 
   can be written as 
\ba\label{eq:vort_sp1aa}
|\nabla A|^2=  \left( \rho |\bm{v}|^2+ \left| \nabla \sqrt{\rho} \right|^2 \right), 
\ea
where the  transformation $A=\sqrt{\rho} e^{i \phi}$ yields $\rho=|A|^2$ and the velocity $\bm{v}=\nabla \phi$. 
Here, following, e.g., the exposition of~\cite{bradley2012energy}, the first and second terms represent the density of the kinetic energy ($E_\text{ke}$) and the quantum pressure ($E^{q}$), respectively, where 
the energies are given by 
\ba\label{eq:vort_sp1bb}
E_\text{ke}=\int n|\bm{v}|^2 d^2 r,~~~E^q=\int |\nabla \sqrt{\rho}|^2 d^2 r. 
\ea
The velocity vector $\bm{v}$ now can be written as a sum over a solenoidal part (incompressible) $\bm{v}^\text{ic} $ and an irrotational (compressible) part $\bm{v}^\text{c}$ as
\ba\label{eq:vort_sp1}
\bm{v}=\bm{v}^\text{ic}+\bm{v}^\text{c}, 
\ea
such that $\dive{\bm{v}^\text{ic}} = 0$ and $\curl{\bm{v}^\text{c}} = 0$. 

 The incompressible and compressible kinetic energies are then \cite{Mithun2021decay}
 \ba\label{eq:vort_sp2}
E^\text{ic,c} = \int d^2 r |\sqrt{n} \bm{v}^\text{ic,c}(\bm{r})|^2. 
\ea

We additionally find the vortex spectra of the system \cite{Eyink_2006,RevModPhys.71.S383,kraichnan1980two,Numasato_2010,Mithun2021decay}. 
In $k$-space ($k$-wave vector), the total incompressible and compressible kinetic energy $E_{ke}^\text{ic,c}$ 
is represented by
 \ba\label{eq:vort_sp3}
 E^\text{ic,c}(k)= k\sum_{j=x,y}\int_0^{2\pi} d\phi_k |\mathcal{F}_j(\bm{k})^\text{ic,c}|^2,  
\ea
where $\mathcal{F}_j(\bm{k})$ is the Fourier transform of $\sqrt{n} u_j$ of the $j$-th component of $\bm{u}= (u_x, u_y)$ and ($k=\sqrt{k_x^2+k_y^2}$, $\phi_k$) represents polar coordinates. 
Through the above Fourier space diagnostics, we can assess
the transfer of energy between different wavenumbers, as well
as the fractions of energy that pertain primarily to the vortical excitations
(associated with the incompressible part) and ones associated
with sound waves (the compressible part), analogously to earlier works such as~\cite{bradley2012energy,Mithun2021decay}.

 \begin{figure*}[!hbtp]
\includegraphics[width = \textwidth]{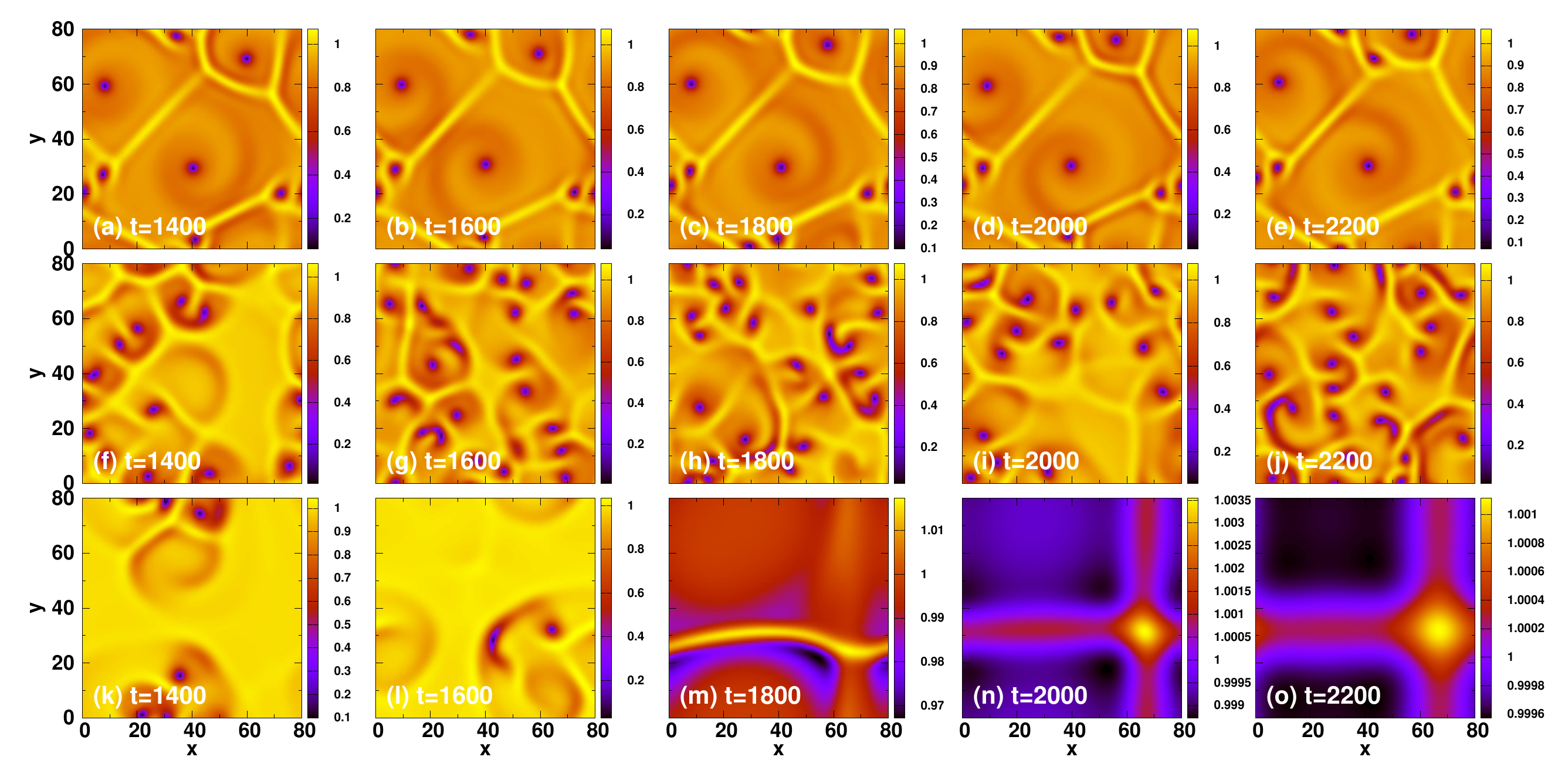}
\caption{ Snapshots of the field amplitude $|A|$ for 
drive strengths (a-e) $A_0=2.0$, (f-j) $A_0=3.6$, and (k-o)  $A_0=4.8$. The other parameters are: repetition time interval $T=10$, and
initial perturbation time $T_0=1200$, for CGL model parameters
$\alpha=0.7$ and $\beta=-0.7$.}
\label{fig:fig2}
\end{figure*}

\begin{figure}[!htbp]
\includegraphics[width = 0.48\textwidth]{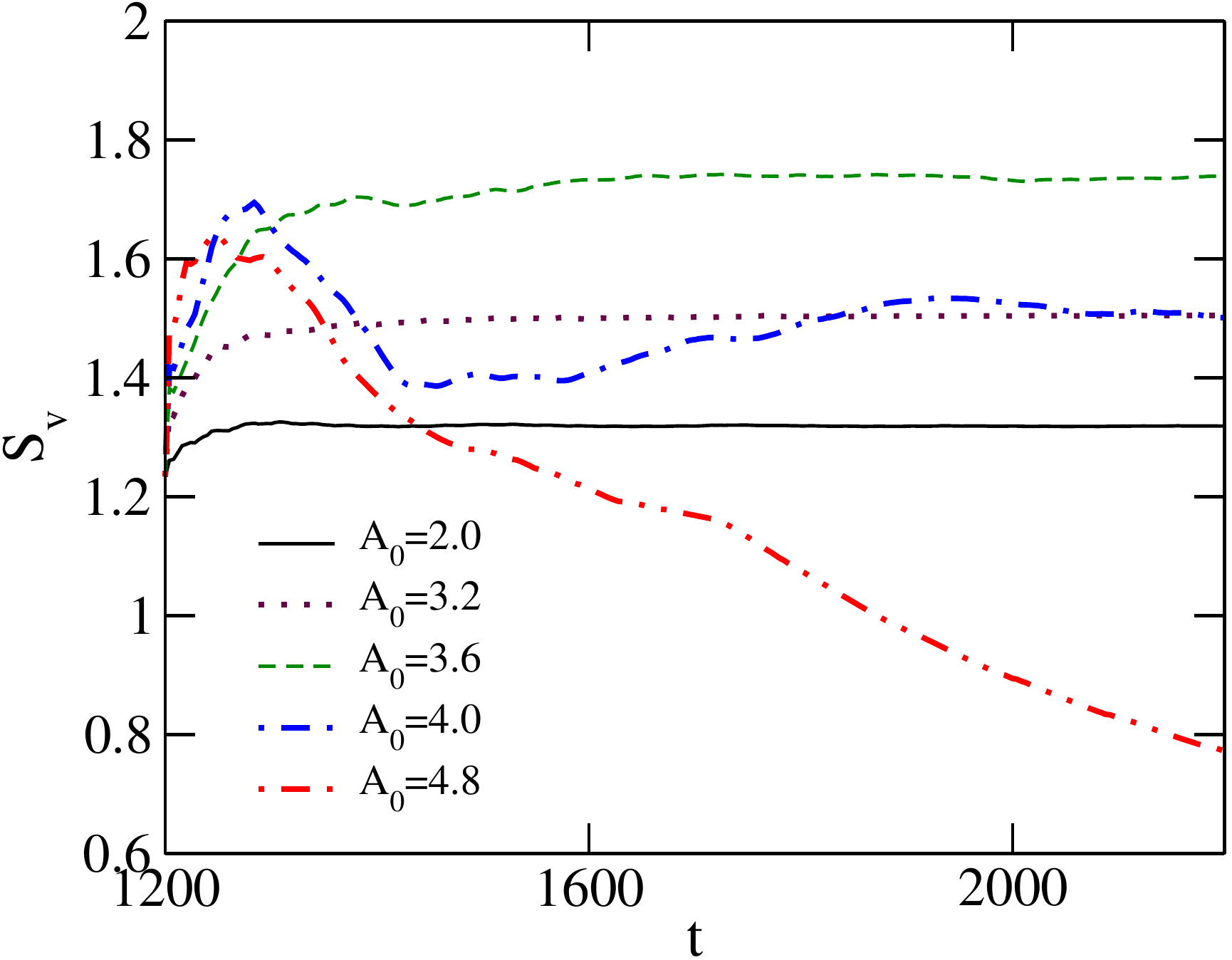} 
\caption{Shannon entropy $S_v$ for $T=10$ for different values of drive amplitude $A_0$. 
}
\label{fig:figS}
\end{figure}

\begin{figure*}[!htbp]
\includegraphics[width = 8.cm]{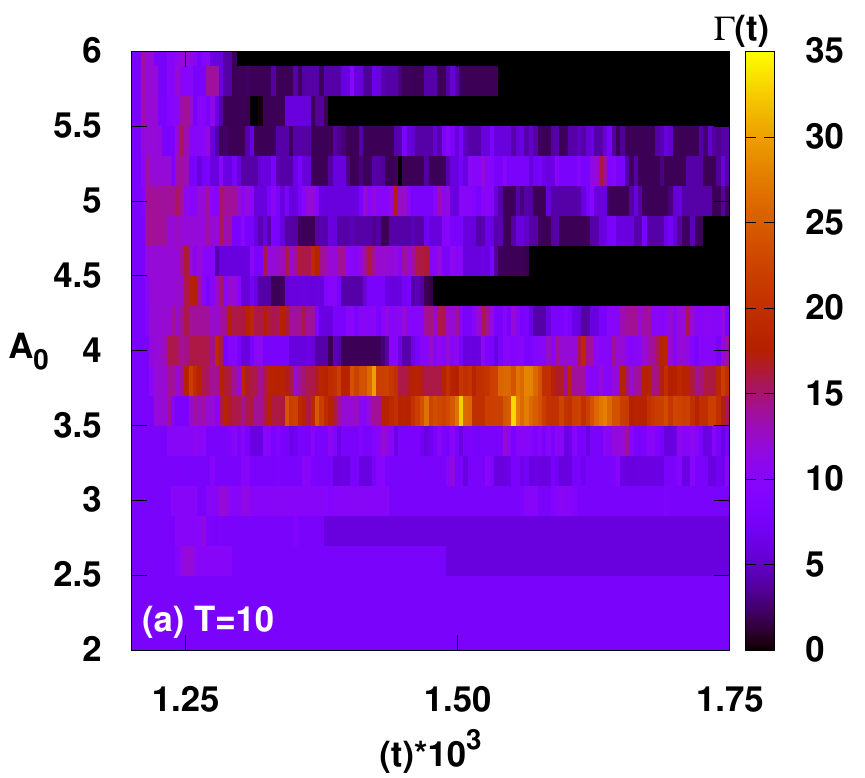}
\includegraphics[width = 8.cm]{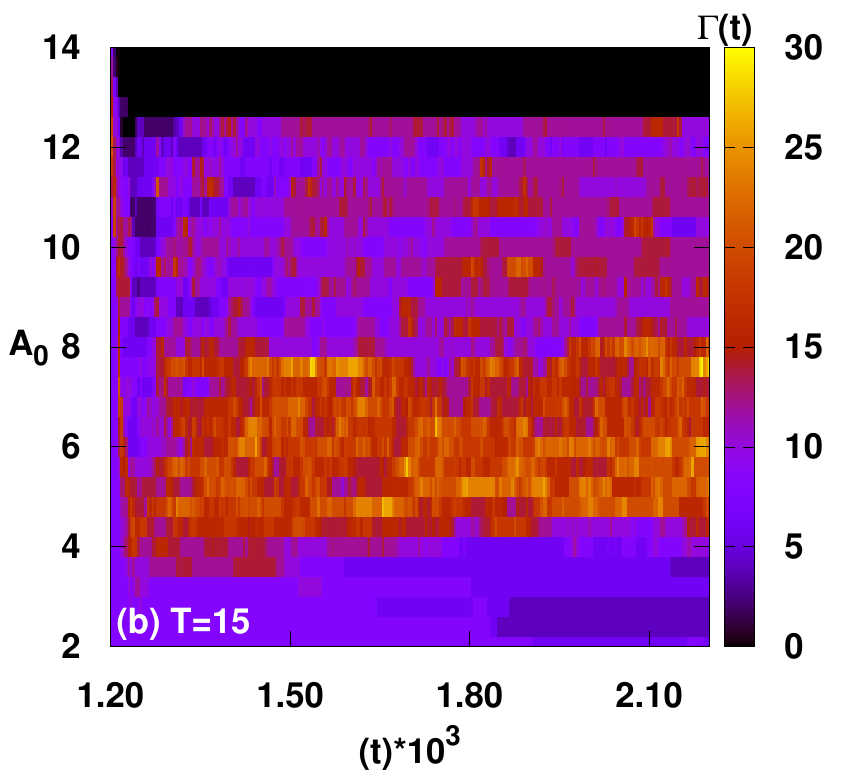}
\caption{Winding number $\Gamma$ for (a) $T=10$ and (b) $T=15$. The axes here are $A_{0}$ (the drive amplitude) and time $t$ (measured
from the time of first kick).}
\label{fig:pbc_wt}
\end{figure*}

 \section{Numerical experiments }

We begin the numerical experiments by considering different values of the driving amplitude $A_0$ for a fixed drive periodicity $T$. Fig.~\ref{fig:fig2} shows the field amplitude for $A_0=2,~3.6~\text{and}~4.8$ and the relevant ``kicks'' repeated every $T=10$ time units.
The figure indicates the existence of three different regimes; (i) the system is nearly unaffected by the periodic driving: this is the regime where coherence is fully
preserved and the relevant perturbation is weak. (ii) Nucleation of new vortex pairs and their subsequent annihilations thereof: this is the
intermediate regime where effective decoherence first emerges, substantially 
modifying the ultimate fate of the system, although the latter is still in the state bearing vortices with domain walls separating them. (iii) Finally, a constant density regime emerges: here, the perturbation is strong and consequently it
entirely collapses the system to a different state with no memory of the initial condition. In regime (i), despite the
fact that no new vortices are generated, the imprint of the 
``measurement'' (or more accurately here, perturbation) process 
arises through the rapid emission of spiral waves which can be clearly
observed in Fig.~\ref{fig:fig2}(a-e). Regime (ii) can be seen in
panels (f-l), while regime (iii) is shown in panels (m-o).

We first characterize these observations using the defined observables. Fig.~\ref{fig:figS} shows the Shannon entropy based on the probability distribution of amplitude $|A|$ for different values of $A_0$. The entropy initially increases with increase in $A_0$ due to the high density fluctuations. At larger times, the entropy nearly saturates in regimes (i) and (ii), while it decreases in regime (iii).  
In regime (i), the saturation is rather natural to expect, as the system rapidly relaxes to the previous glassy state. 
Similarly in regime (iii), the state evolves towards a constant
amplitude (of unity), hence naturally the attraction to this
state leads the entropy to a rapid decrease towards $0$. The most
dynamic state is the intermediate one of regime (ii), where the
relaxation is slow and therefore the entropy presents the oscillatory
dynamics observed in Fig.~\ref{fig:figS}.

The measurement based on Shannon entropy provides a cumulative
diagnostic (rather than a distributed one) that clearly highlights the effect of periodic driving on the amplitude.  At the same time, the entropy does not immediately provide information about the vortex configurations (aside perhaps from regime (iii) where the
tendency of the entropy to go to $0$ can lead us to infer the absence
of vorticity). In that vein, we now characterize the dynamics based on the absolute value of the winding number that provides us with a sense
of the vortex creation and annihilation processes occurring in the
system as a result of the drive. We consider two periods $T=10$ and $T=15$. 
The results of both cases shown in Fig.~\ref{fig:pbc_wt} illustrate that as $A_0$ increases, the number of initially generated vortices also increases in accordance with  Fig.~\ref{fig:fig2}. The results of the $T=10$ case show that for lower values of $A_0=(2,2.4)$, the system does not respond to the kicks
by changing its vortical content; once again, this illustrates 
the regime (i) of effective coherence preservation. 

For larger values of kick amplitude, on the other hand, the number of initially generated vortices increases (in the presence of the
perturbations), yet at the same time, the rate of vortex-antivortex annihilations also progressively increases with $A_0$. On the other hand, for $T=15$ the rate of annihilation is decreased as compared to the $T=10$ case. This is due to the fact that the system has more time to relax in between the kicks as $T$ increases. For intermediate values of 
$A_0$, we therefore observe that the time evolution of the system
preserves a dynamic effect where significant (additional) vorticity
is present, as shown by the colorbar of Fig.~\ref{fig:pbc_wt}
and the system has decohered from its original glassy state.
For large values of $A_0$, eventually the vortex annihilation
events take over and the system reaches the stable, locally
attracting equilibrium state of $|A|=1$.

\begin{figure*}[!htbp]
\includegraphics[width = 8.cm]{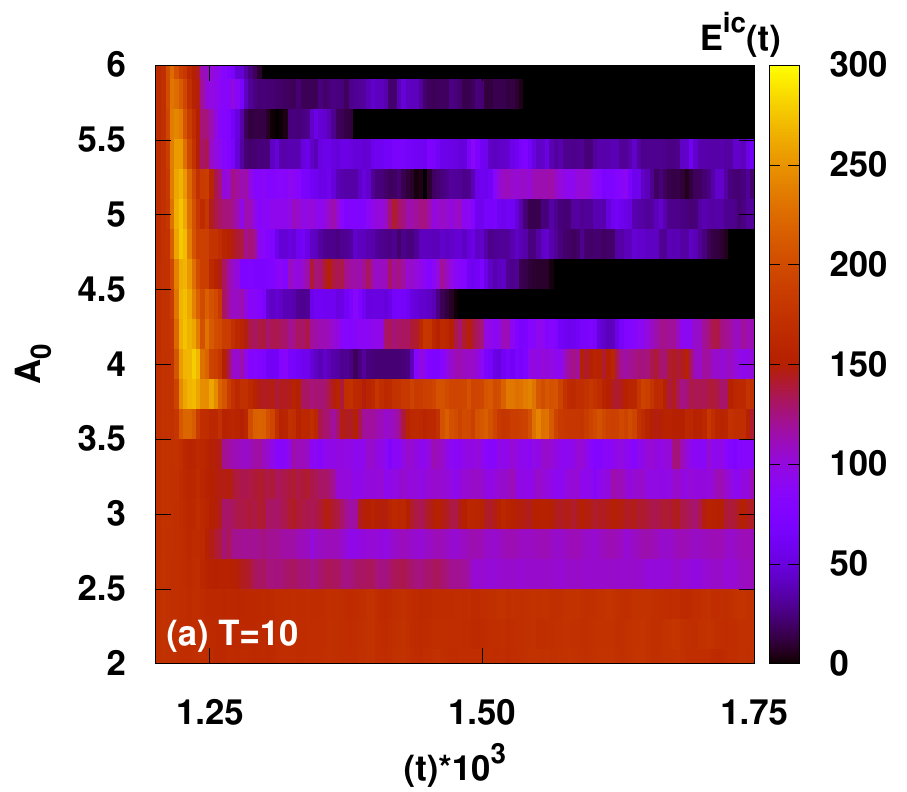}
\includegraphics[width = 8.cm]{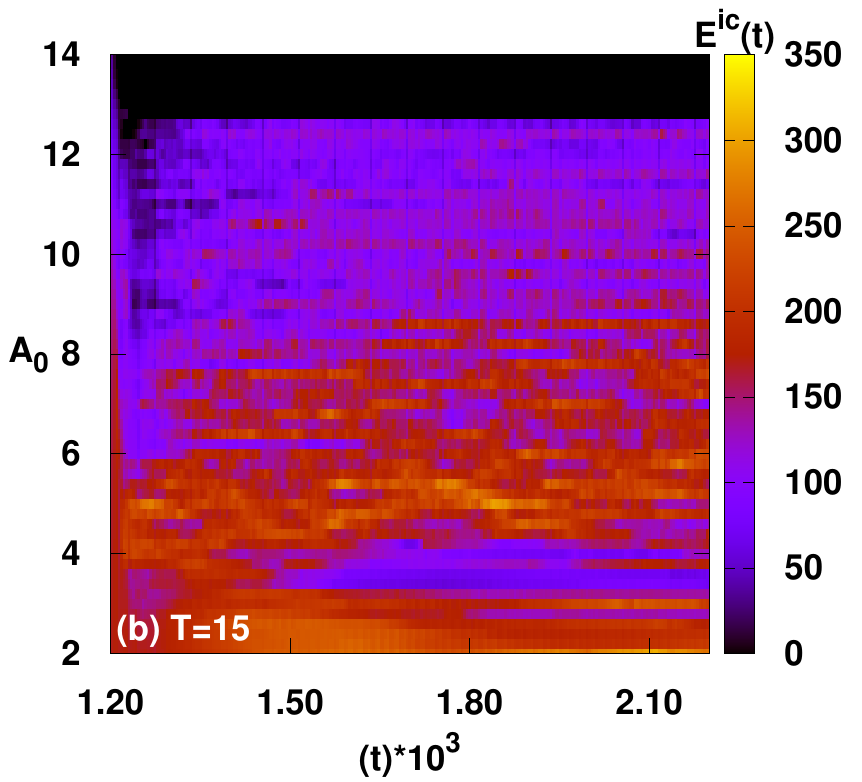}
\caption{ Incompressible energy as a function of time for (a) $T=10$ and (b) $T=15$. Colors represent different values of the
drive strength $A_{0}$. }
\label{fig:EIC}
\end{figure*}
\begin{figure*}[!htbp]
\includegraphics[width = 8.cm]{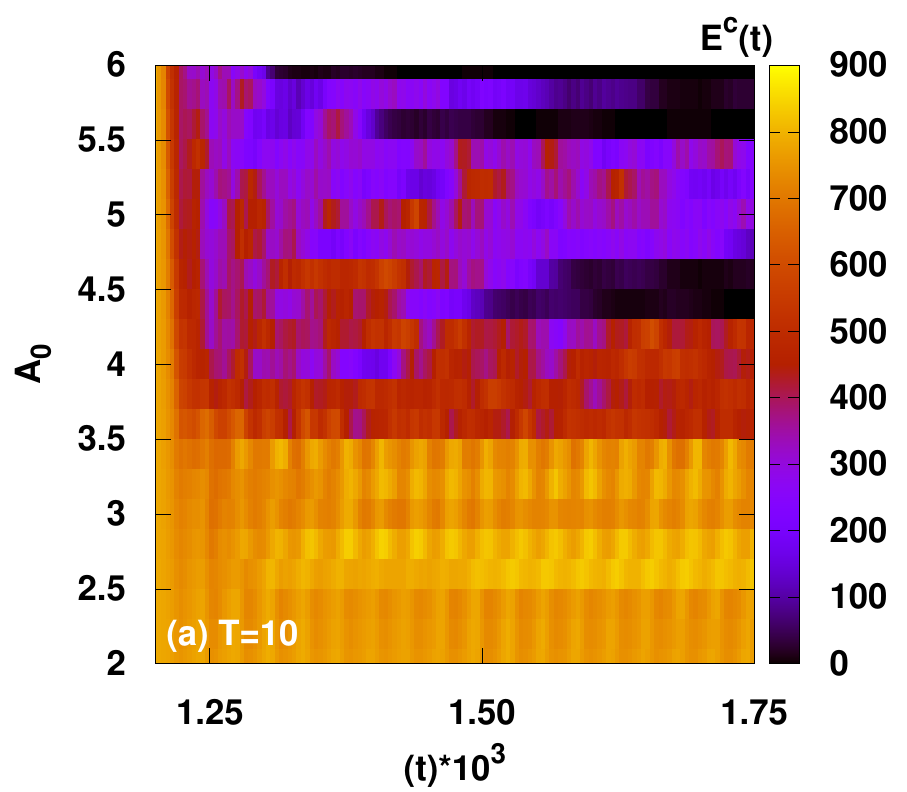}
\includegraphics[width = 8.cm]{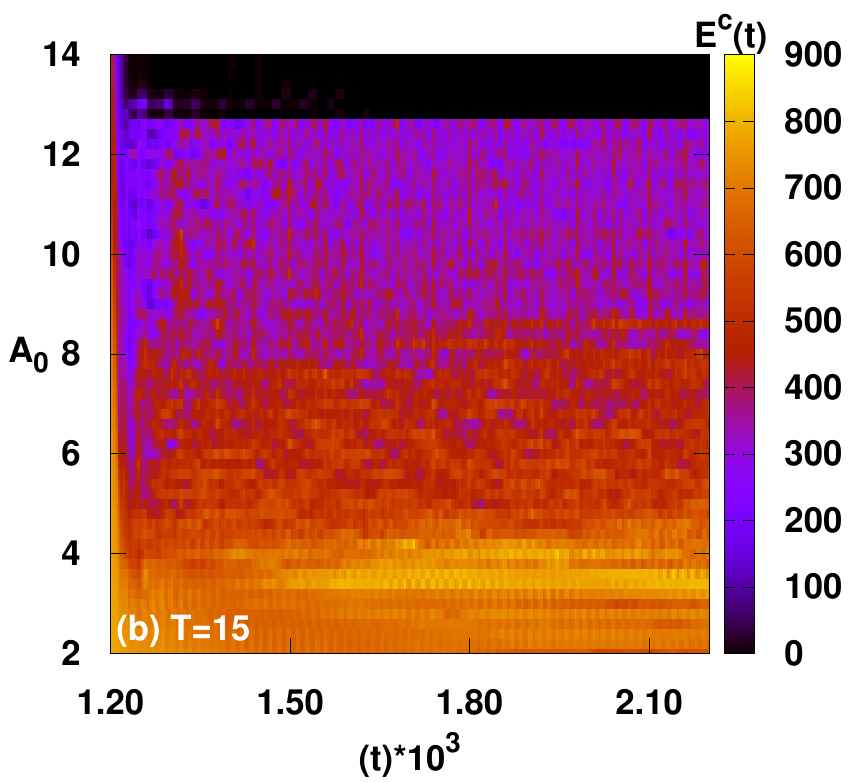}
\caption{Compressible energy as a function of time for (a) $T=10$ and (b) $T=15$. Colors represent different values of the drive
strength $A_{0}$.}
\label{fig:EC}
\end{figure*}

The measured incompressible and compressible energies for  the
cases of $T=10$ and $T=15$  are shown in Fig.~\ref{fig:EIC} and Fig.~\ref{fig:EC}, respectively. As expected, for $T=10$, the incompressible energy is nearly a constant for $A_0=(2,2.4)$ [see Fig.~\ref{fig:EIC}(a)], 
representing the invariance of the vortex configuration. For the range $A_0=2.6-2.8$  the incompressible energy initially decreases and then saturates at later times corresponding to the one vortex pair reduction shown in Fig.~\ref{fig:pbc_wt}(a).
The increase in the rate of annihilation is well reflected in  Fig.~\ref{fig:EIC}(a) for higher $A_0$, where incompressible energy decreases. Similar results are seen in Fig.~\ref{fig:EIC}(b) for $T=15$, although the larger relaxation time induces the associated phenomenology
at larger values of $A_0$.
For higher values of $A_0$, it is clear that the incompressible 
energy is fluctuating in a way that parallels the winding
number and reflects the potential emergence (for intermediate
values of $A_0$) and disappearance (for higher values of $A_0$)
of vortex-antivortex pairs through the relevant creation and
annihilation processes.

\begin{figure*}[!htbp]
\includegraphics[width = 5.8cm]{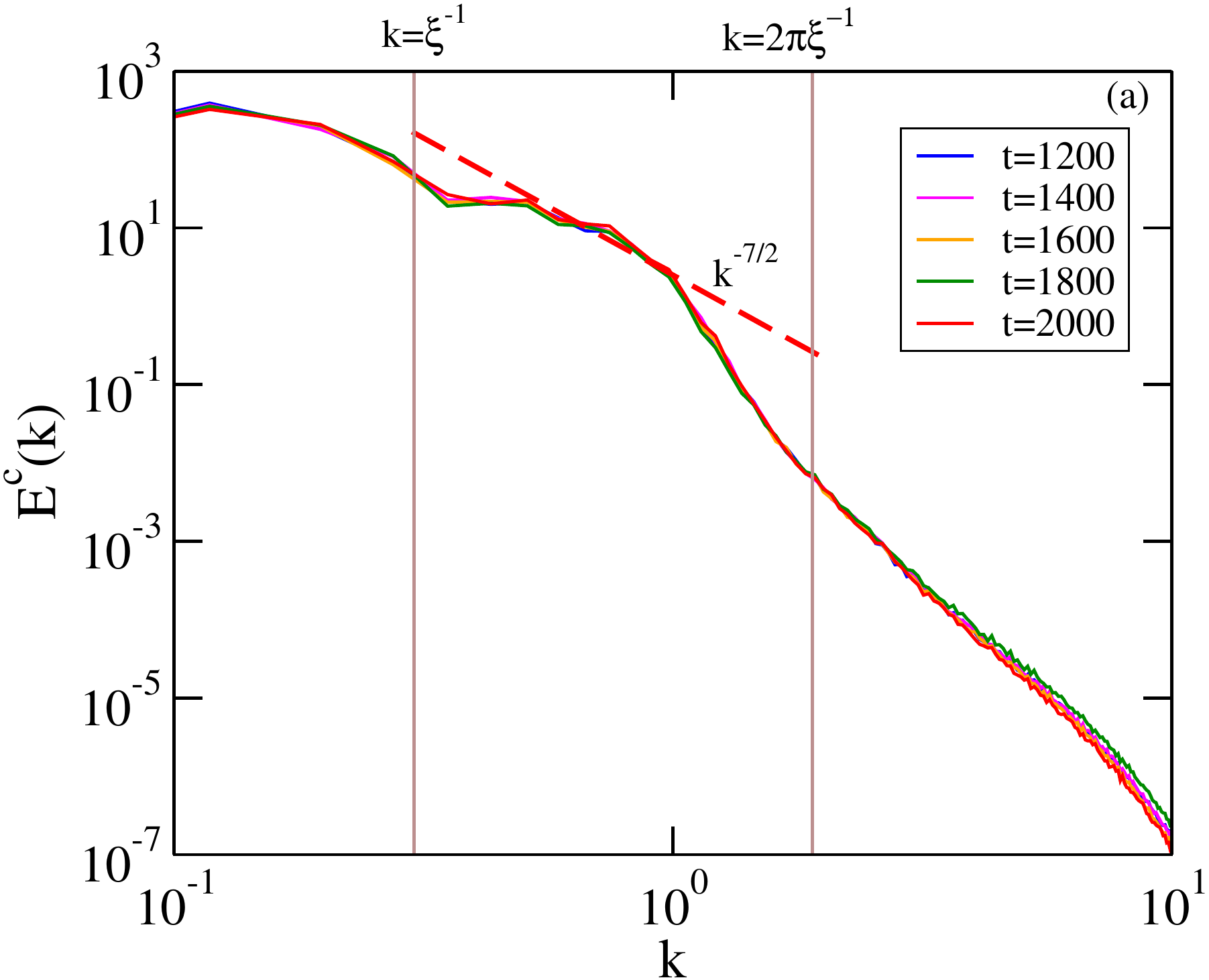}
\includegraphics[width = 5.8cm]{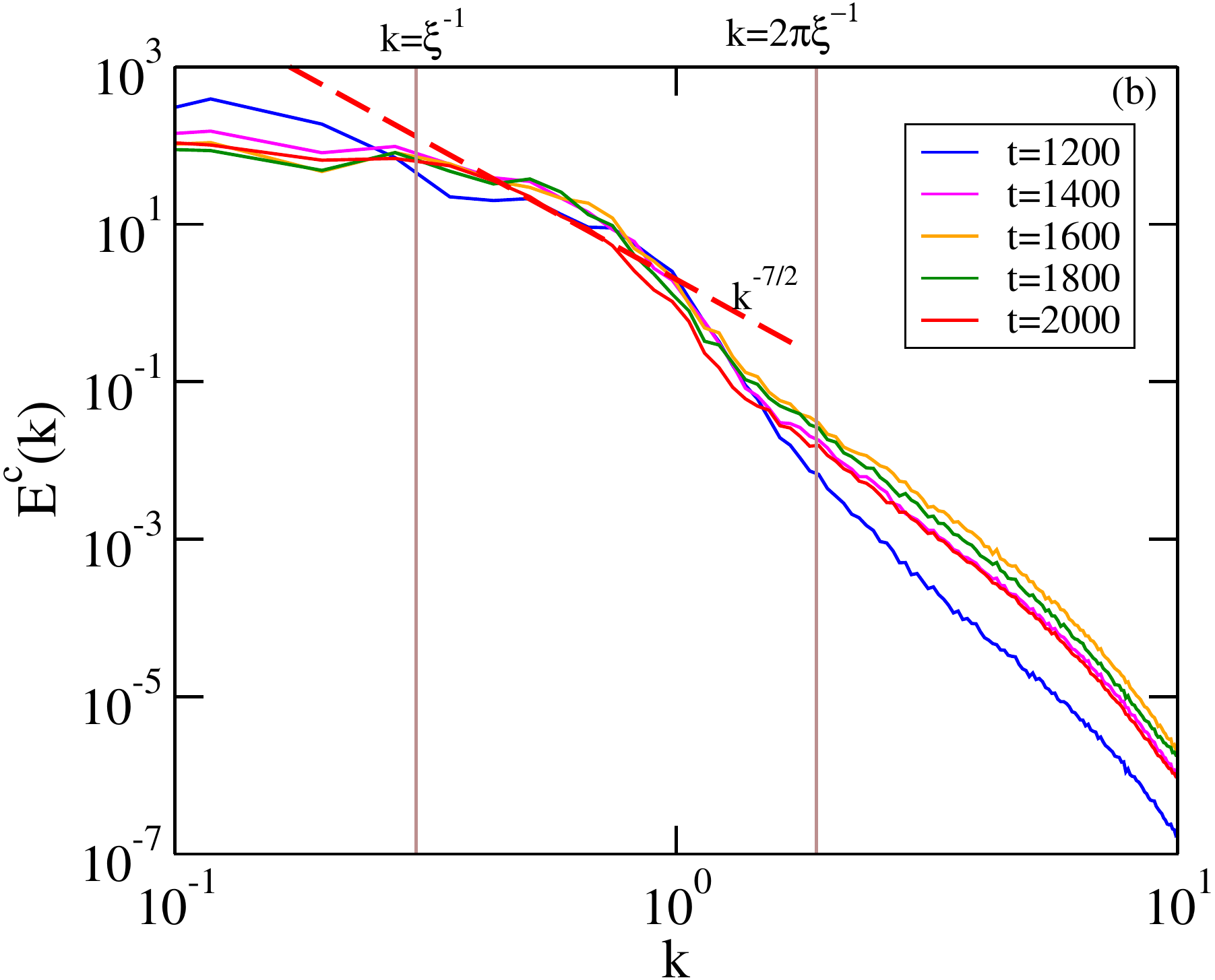}
\includegraphics[width = 5.8cm]{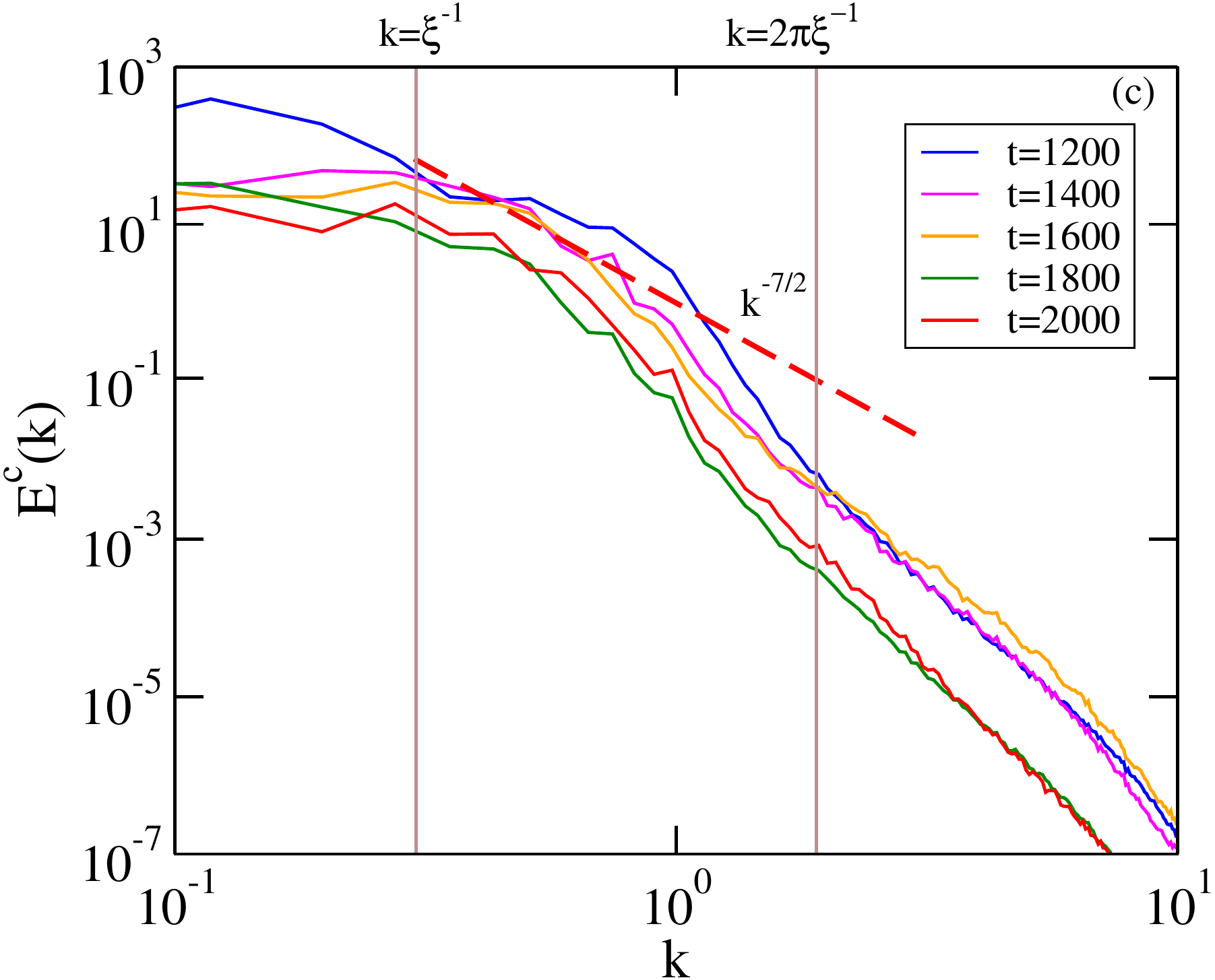}
\includegraphics[width = 5.8cm]{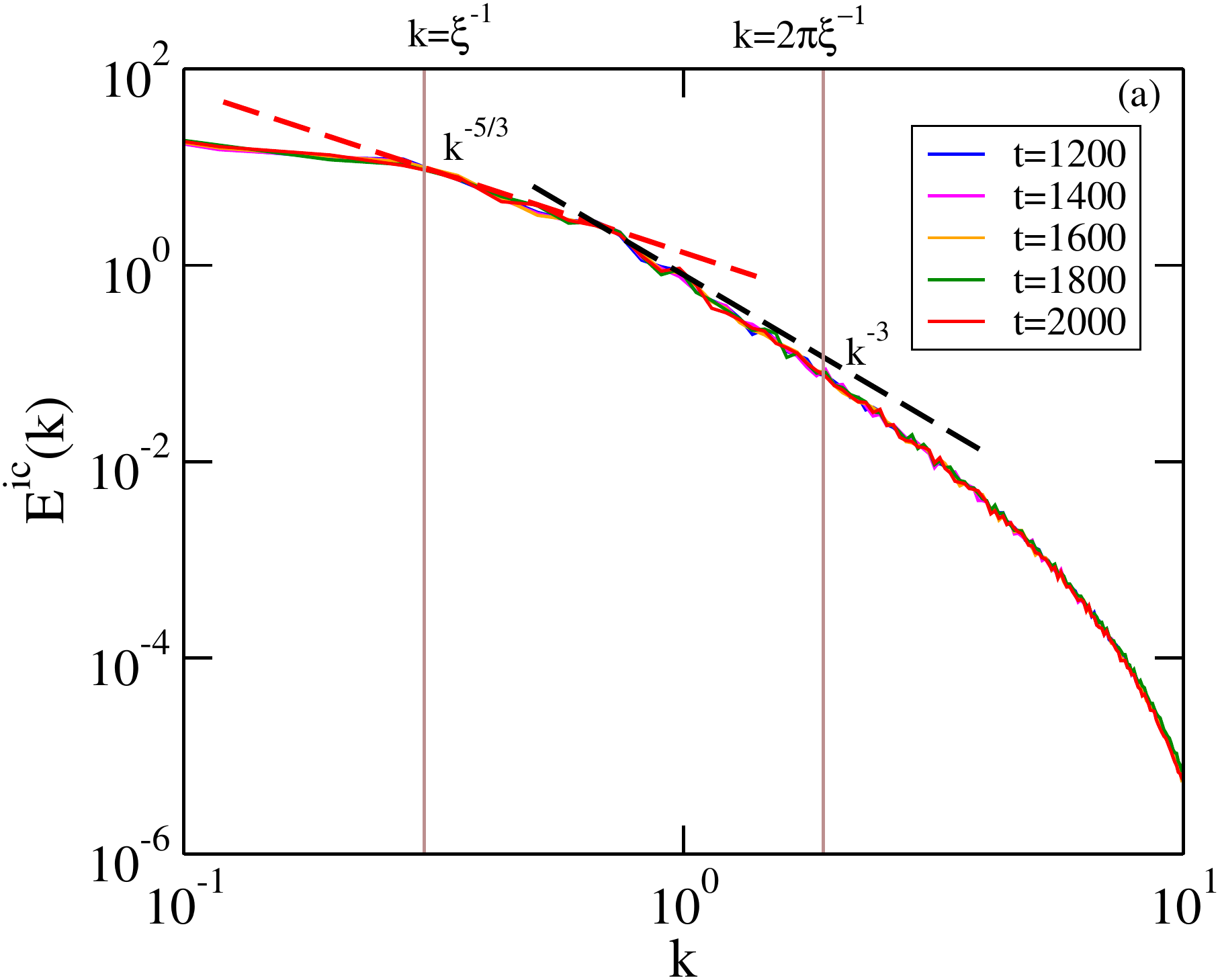}
\includegraphics[width = 5.8cm]{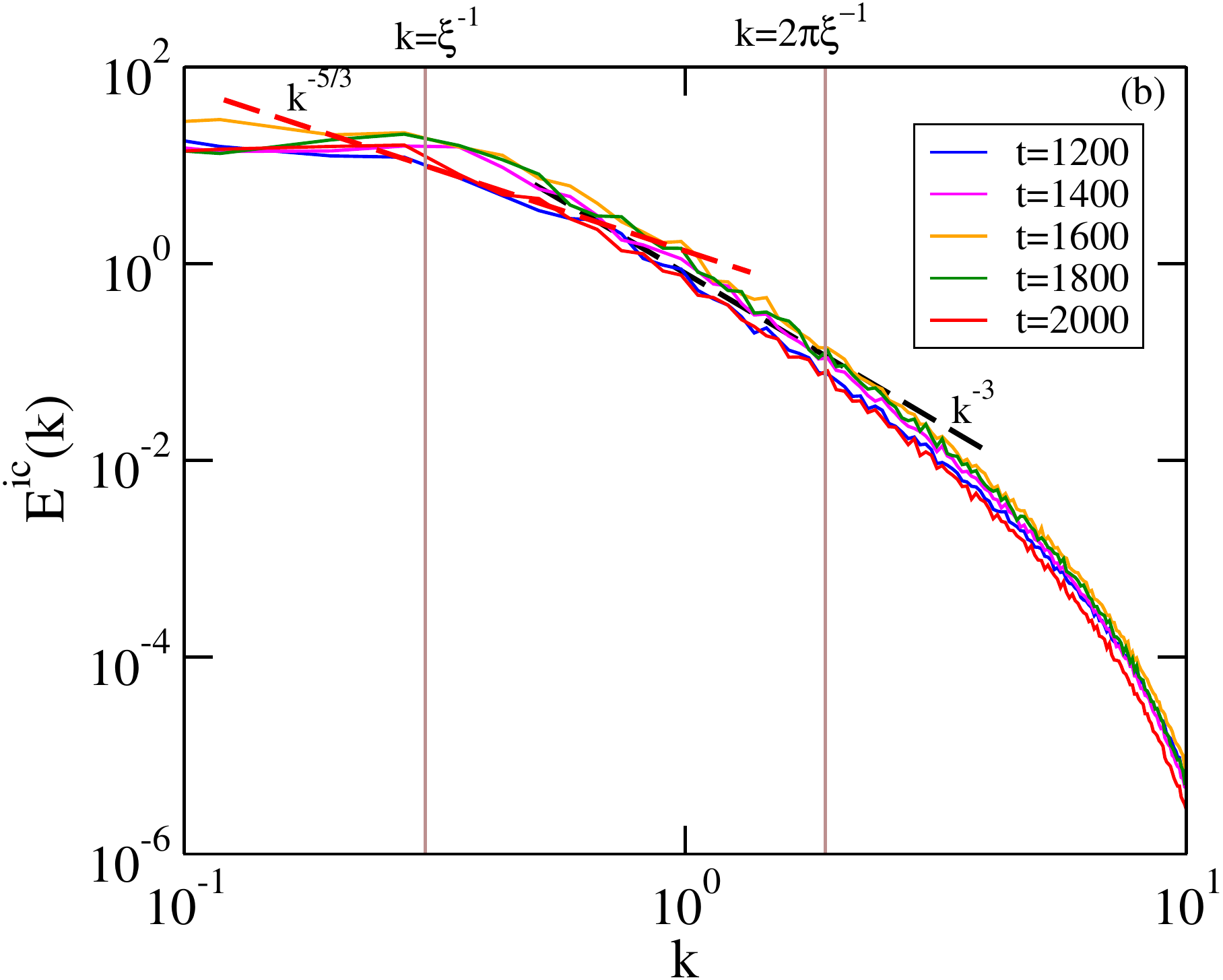}
\includegraphics[width = 5.8cm]{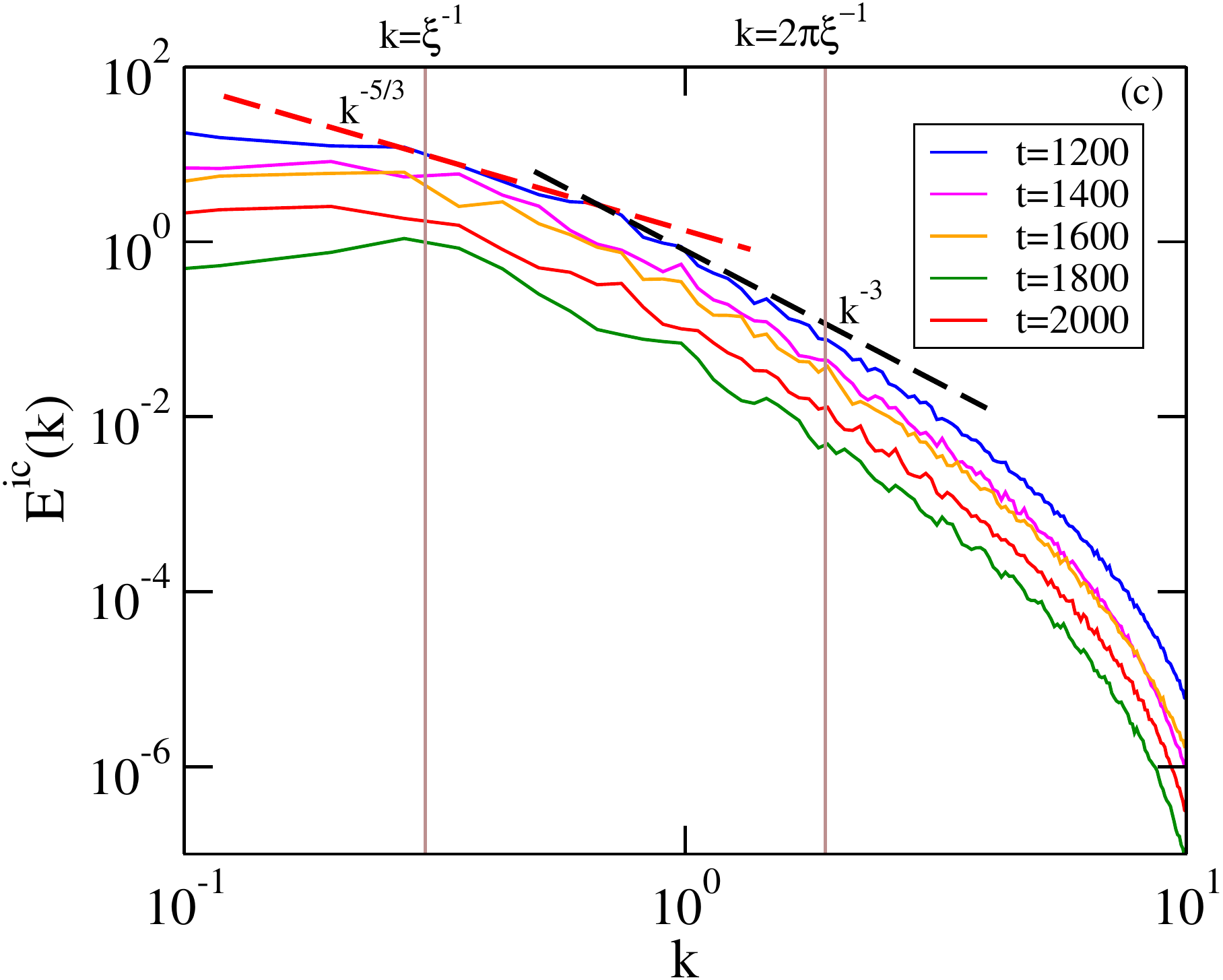}
\caption{(Top panel) Compressible energy spectra for (a) $A_0=2$, (b) $A_0=3.6$ and (a) $A_0=4.8$. (Bottom panel) The corresponding incompressible energy spectra for (a) $A_0=2$, (b) $A_0=3.6$ and (a) $A_0=4.8$. The vertical lines (from left to right)
represent $k=1/\xi$ and $k=2 \pi/\xi$, where $\xi$ is
the so-called healing length associated with the size of individual
vortices. The other parameters are $T=10$ and $\alpha=0.7$ and $\beta=-0.7$.}
\label{fig:T10spectra}
 \end{figure*}

 
We next analyze  both the compressible $E^\text{c}(k)$ and incompressible $E^\text{ic}(k)$ energy spectra, of the system collected at different evolution times. This is typically done in order
to appreciate the energy exchanges in a system and the
possible connection of the associated mechanisms with potential
turbulent or coherent-structure (such as vortex) induced dynamics~\cite{bradley2012energy}. The energy distribution from smaller scale (corresponding to the vortex core size) to larger scale (system size) can be inferred from such spectra. The upper and lower panels in Fig.~\ref{fig:T10spectra} show the compressible and incompressible energy spectra for three different values of $A_0~(=2.0,3.6,4.8)$ for fixed $T=10$.  
The compressible vortex spectra shown in the upper panel of Fig.~\ref{fig:T10spectra} do not exhibit any convincing scaling laws. However, at the same time we note that for $1/\xi < k < 2\pi/\xi$, where $\xi$ represents the numerically measured healing length (vortex core radius), spectra are closer to the reference line drawn for $k^{-7/2}$. 
This power-law is a characteristic of a superfluid turbulence corresponding to the sound wave equilibrium,  as discussed, e.g., in~\cite{Mithun2021decay,PhysRevA.86.053621}.  The decrease in compressible energy with increase in time for higher $A_0$ is due to the loss of small amplitude fluctuations as shown in Fig.~\ref{fig:fig2}.

The incompressible energy spectra of the same case are depicted in the bottom panel of Fig.~\ref{fig:T10spectra}.  The spectra of different values of $A_0$ show the existence of $k^{-3}$-power law, associated with individual vortex cores~\cite{bradley2012energy},
for sufficiently large $k$. On the other hand,
interestingly and somewhat unexpectedly, there exists a lower interval
of wavenumbers where our observations are close to a Kolmogorov $k^{-5/3}$ power-law that is typically associated in 2D settings
with the inverse energy cascade as observed, e.g.,  in 
superfluid vortex turbulence \cite{bradley2012energy}. We conjecture that the existence of $k^{-5/3}$ power-law, even for smaller $A_0$, is due to the irregular arrangement and slow motion of vortices, although this is
clearly a topic that merits further investigation. The downward shift of vortex spectra at larger times for larger values of $A_0$ is due to the decrease in the vortex numbers. We find similar results for the $T=15$ case as well (results omitted for brevity).

In this
section we demonstrated the effect of a periodic drive on the defect dynamics of the CGL model. Our results suggest that there
are cascading processes even in the weaker amplitude
cases, where the configuration at the level of its density profile
appears to preserve its coherence and glassy state structure.
For higher values of $A_0$, it is apparent (see, e.g., the middle
and right panels of Fig.~\ref{fig:T10spectra}) that there are 
stronger energy exchange mechanisms at work, both ones mediated
via the substantial additional compressible (sound) energy and
ones realized via the vortex-antivortex creation (hence
increase of incompressible energy) or annihilation (hence 
conversion of the incompressible energy into sound waves). 
In this case of larger $A_0$, these processes result in the
loss of memory and effective (nonlinearity-driven) decoherence
of the dynamics.

%
\section{Conclusions and Future Challenges}
One of the motivations for this work has been to explore a classical analog
of quantum effects such as  decoherence via suitable nonlinear 
classical field theoretic examples.
The driven complex Ginzburg-Landau CGL) equation in 2D
offers an intriguing playground for the exploration of these effects as it possesses certain instabilities 
and coherent topological structures such as vortices and domain walls, and features 
wide parametric regimes of metastable, long-lived states such as the vortex
glasses utilized here~\cite{bohr,PhysRevE.65.016122}. 
In the periodically driven version of this dissipative nonlinear system that we
considered here, 
we have aimed to establish a phase diagram that offers a glimpse into a classical (nonlinearity-induced) analogue of the notion of decoherence. In particular, the axis of variation of the period of the drive
$T$ is an analog of the ``measurement rate", while the 
amplitude axis $A_0$ as an analog of the ``mixing rate". As we have shown, we
indeed numerically find a demarcation curve between the examined 
cases, with
a parametric region representing decoherence and another preserving
the system's statistical memory; the intermediate regime between these two
features the most dynamical environment where vortex creation-annihilation
events most dramatically and persistently take place. 

We believe that this study poses a number of challenges that are worth pursuing in the future work.  
Clearly, the form, in space and time, of the external field we have used here has many potential variations
beyond the spatially
uniform, time periodic one used here, analogous to various measurement protocols employed in quantum technologies. 
A systematic comparison of quantum systems with classical nonlinear ones as regards quantum features such 
as entanglement and decoherence is called for, including both dissipative and non-dissipative (open and closed) 
cases. 
The Gross-Pitaevskii model and its many-body variants~\cite{lode1} 
could represent an ideal framework for exploring the analogies and differences between classical nonlinear 
and genuinely quantum systems \cite{PhysRevA.102.033301}. Moreover, 
the transverse field Ising and similar models 
provide examples where coherence times can be equated with correlation lengths in equivalent 
higher-dimensional classical models \cite{Stinchcombe_1973}. Finally, in the present report, we have
explored issues pertaining to driving in connection to coherence vs. decoherence and memory vs. memory loss, 
but have not examined aspects pertaining to entanglement. The latter may be an especially interesting topic 
(along with its similarities and differences to wave interactions) for future study.

\section{Acknowledgements}
This material is based upon work supported by the US
National Science Foundation under Grants No. PHY-2110030 and DMS-1809074 (PGK).  The work at Los Alamos National Laboratory was carried out under the auspices of the U.S. DOE and NNSA under Contract No. DEAC52-06NA25396.
\bibliographystyle{apsrev4}
\let\itshape\upshape
\normalem
\bibliography{reference1}




\end{document}